\documentclass[a4paper,nofootinbib,superscriptaddress]{revtex4}

\usepackage{color}
\usepackage{amsmath,amsfonts,amssymb,amsthm}
\usepackage{graphicx}
\usepackage{fullpage}
\PassOptionsToPackage{caption=false}{subfig}
\usepackage{subfig}

\usepackage{epstopdf} % to include .eps graphics files with pdfLaTeX
\usepackage{bm}  % Define \bm{} to use bold math fonts

\usepackage[pdfpagelabels,pdftex,bookmarks,breaklinks]{hyperref}
\definecolor{darkblue}{RGB}{0,0,127} % choose colors
\definecolor{darkgreen}{RGB}{0,150,0}
\hypersetup{colorlinks, linkcolor=darkblue, citecolor=darkgreen, filecolor=red, urlcolor=blue}

\usepackage{layout}

\def\Z{\mathbb{Z}}

\newcommand{\Eref}[1]{Eq.~(\ref{#1})}
\newcommand{\Sref}[1]{Sec.~\ref{#1}}
\newcommand{\Fref}[1]{Fig.~\ref{#1}}

\def\th{^{\rm th}}
\def\nd{^{\rm nd}}

\newcommand{\ket}[1]{|{#1}\rangle}

\newcommand{\ketbra}[2]{|{#1}\rangle\!\langle{#2}|}

\newcommand{\proj}[1]{\ketbra{#1}{#1}}

\begin{document}

\title{Generalized Color Codes Supporting Non-Abelian Anyons}

\author{Courtney G.\ Brell}
\email{courtney.brell@gmail.com}
\affiliation{Centre for Engineered Quantum Systems, School of Physics, The University of Sydney, Sydney, Australia}
\affiliation{Institut f\"ur Theoretische Physik, Leibniz Universit\"at Hannover, Appelstra\ss e 2, 30167 Hannover, Germany}

%\date{\today}

\begin{abstract}
We propose a generalization of the color codes based on finite groups $G$. For non-abelian groups, the resulting model supports non-abelian anyonic quasiparticles and topological order. We examine the properties of these models such as their relationship to Kitaev quantum double models, quasiparticle spectrum, and boundary structure.
\end{abstract}

\maketitle

%------------------------------------------------------------------------------------------------------------%
\section{Introduction}\label{S:intro}
%------------------------------------------------------------------------------------------------------------%

Topological codes are a promising avenue to achieve robust quantum memories~\cite{Dennis2002} or implement fault-tolerant quantum computation~\cite{KitaevTC97}. These codes have locality properties that are both advantageous from the perspective of implementation, and give robustness against realistic noise models. Topologically ordered systems can be used for information processing in a number of ways, notably by code deformation~\cite{Bombin2009a, Bombin2011b} or by the braiding of quasiparticle excitations~\cite{Nayak2008}. The latter approach is available only for particular types of topologically ordered systems with non-abelian anyonic excitations.

The color codes~\cite{Bombin2006,Bombin2007} are a family of topological codes with abelian anyonic excitations. They may be used to perform computation by code deformation, but are particularly notable for having a large class of transversal gates~\cite{Bombin2006}, giving rise to high fault-tolerance thresholds~\cite{Landahl2011}. They are also related to many other interesting families of codes such as the toric codes~\cite{KitaevTC97}, topological subsystem codes~\cite{Bombin2009, Bombin2010}, higher-dimensional color codes~\cite{Bombin2007a}, and gauge color codes~\cite{Bombin2013}. Small examples of color codes have also been demonstrated and manipulated in the laboratory~\cite{Nigg2014}.

While the color codes have interesting properties and are related to many other interesting models, the ability to support non-abelian excitations is one feature they lack. Here, we present a generalization of the color code to arbitrary finite group $G$ (such that the standard color code corresponds to the group $\Z_2$). This is motivated in analogy to the generalization of the toric code to the quantum double models~\cite{KitaevTC97}. These generalized color codes support non-abelian anyons for non-abelian groups $G$ and so may in general be used for topological quantum computation by braiding these quasiparticles.

We also study some notable properties of these generalized color codes. Particularly, we demonstrate an equivalence between the generalized color codes and the quantum double models that allows us to easily determine the quasiparticle content of these color codes. We further discuss the structure of the boundaries of these models among other properties.

The layout of the paper is as follows: in \Sref{s:color}, we will review the qubit color code model and introduce the generalized color code. We will prove the relation between these models and the quantum double models in \Sref{s:qdmap}. Following this, in section \ref{s:properties} we will explore the properties of the generalized color codes, before concluding remarks in \Sref{s:discussion}.

%------------------------------------------------------------------------------------------------------------%

\newcommand{\siteN}[4]{
\begin{array}{cc}#1&#2\\#4&#3
\end{array}
}

\section{Qubit Color Codes and $G$-Color Codes}\label{s:color}

\begin{figure}
		\centering
		\subfloat[]{\label{f:666lattice}
		\includegraphics{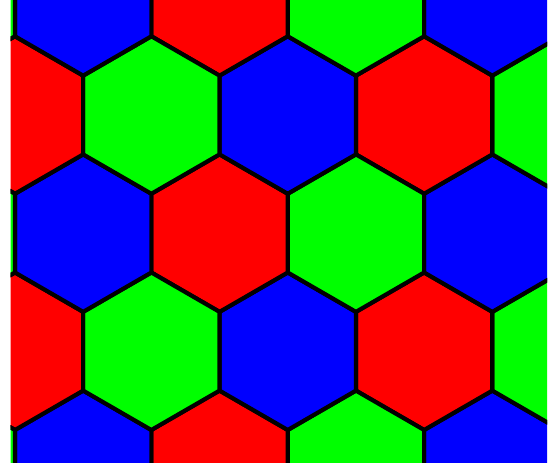}
		}
		\hspace{1cm}
		\subfloat[]{\label{f:488lattice}
		\includegraphics{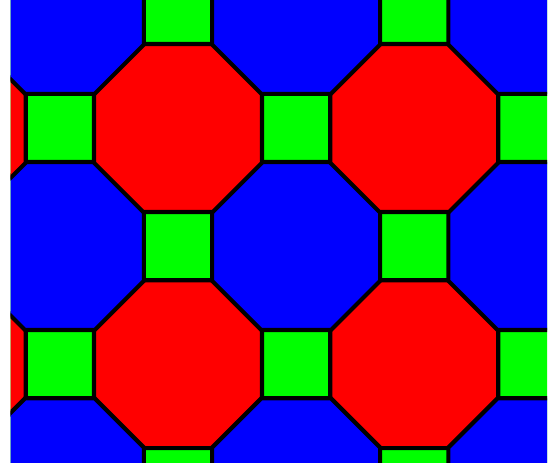}
		}
		\caption{Two examples of 2-colexes. (a) The 6.6.6 (honeycomb) lattice. (b) The 4.8.8 lattice.}
		\label{f:2colexes}
		\end{figure}

		The qubit color code~\cite{Bombin2006} is defined on a trivalent lattice whose plaquettes are 3-colorable (see for example \Fref{f:2colexes}). Such lattices are called 2-colexes~\cite{Bombin2007a}. Qubits are placed on vertices of the 2-colex, and each plaquette has two associated projectors, defined as
	\begin{align}
		S_p^X &=\frac{1}{2}\left(1+\prod_{i\in p}X(i)\right)\\
		S_p^Z &=\frac{1}{2}\left(1+\prod_{i\in p}Z(i)\right)
	\end{align}
	such that $i\in p$ runs over vertices bounding the plaquette $p$, and where $X(i)$ and $Z(i)$ are the Pauli matrices acting on vertex $i$. 2-colexes are always bipartite, which means that all cycles of the lattice are even in length, and so $S_p^X$ and $S_p^Z$ commute with each other for the same $p$. It can also be seen that $[S_p^X,S_{p'}^Z]=0\;\forall p,p'$.  The model is then defined by the Hamiltonian
	\begin{align}
		H = -\sum_p\left[S_p^X+S_p^Z\right]
	\end{align}
	such that the ground space of the code is the common $+1$ eigenspace of each of the $S_p$ operators. We will therefore refer to the $S^X$ and $S^Z$ as $X$- and $Z$-type stabilizers.
	
%------------------------------------------------------------------------------------------------------------%
		
%------------------------------------------------------------------------------------------------------------%

	Plaquettes of the 2-colex are colored red, green, and blue, and similarly each link has an associated color, such that it connects two plaquettes of its own color. The elementary excitations of this model can be thought of as (abelian) anyonic quasiparticles, corresponding to stabilizer operators that are frustrated. Anyon species can be labelled by the color of the plaquette on which they live and the type of stabilizer they frustrate (i.e. $X$ or $Z$), or can be a composite of these generating anyons.
	
	On a torus the qubit color codes are $2^4$-fold degenerate, and alternative boundary conditions for these codes are discussed further in \Sref{s:boundary}. Logical operators in these codes consist of homologically non-trivial strings of $X$ or $Z$ operators running along edges of a particular color, and can branch if strings along edges of all three colors meet. A $X$ string running along red links will anticommute with a $Z$ string running along blue links, for example, and these string operators will form a Pauli algebra acting on the degenerate codespace.	
	
	We can think of the qubit color code as being based on the group $\Z_2$. The qubits on each link have a natural basis labelled by elements of $\Z_2 = \{0,1\}$ (with $0$ the identity element) and we can consider the $X$ operator to act on these states as group multiplication by the $1$ element, i.e.
	\begin{align}
		X \ket{g} = \ket{g\oplus 1}
	\end{align}
	where of course addition modulo 2 is the relevant group multiplication operation for this group. The $X$ operator can also be labelled with a group element superscript such that
	\begin{align}
		X^h\ket{g} = \ket{g\oplus h}
	\end{align}
	and see that $X^1=X$, $X^0=I$.	
	
	Let us also introduce operators $T^g = \proj{g}$ for each $g\in\Z_2$. This allows us to write
	\begin{align}
		Z = T^0-T^1
	\end{align}
	
	In particular, this allows us to rewrite
	\begin{align}
		S_p^X &=\frac{1}{2}\sum_{g\in \Z_2}\prod_{i\in p}X(i)^g\\
		S_p^Z &=\sum_{\bigoplus g_i=0}\prod_{i\in p}T(i)^{g_i}
	\end{align}
	where we use notation such that $\bigoplus g_i=0$ runs over all sets of $g_1,g_2,\ldots,g_n$ such that $g_1\oplus g_2\oplus\cdots\oplus g_n = 0$.

	When written in this form, these operators bear a close resemblance to the $A_v$ and $B_p$ projectors used to define the quantum double models~\cite{KitaevTC97} (in this case for the group $\Z_2$).

	\subsection{$G$-color codes}
		
		Given our interpretation of the qubit color code in terms of the group structure of $\Z_2$, we now define a generalized color code. The desirable characteristics of this code are that its Hamiltonian can be expressed as a (negative) sum of commuting local projectors, and that its excitation spectra includes non-abelian anyons for sufficiently complicated group $G$. The models we present will satisfy both of these conditions. It is worth noting that our models defined for the (abelian) cyclic groups $\Z_d$ will correspond precisely to the known generalizations of color codes to higher-dimensional spins~\cite{Sarvepalli2010}.
		
		A $G$-color code is defined uniquely by the following set of data:
		\begin{itemize}
		\item A finite group $G$
		\item A 2-colex $L$
		\item A parity function $s_i=\pm 1$ at each site $i$ of $L$
		\item A choice of privileged color (conventionally red), and a choice of ``clockwise'' and ``anti-clockwise'' for the remaining colors (conventionally green and blue respectively)
		\end{itemize}

		Here we will begin by defining the model for the special case $s_i=+1\;\;\forall i$, and subsequently describe how to relate models with different parity functions $s_i$. To each vertex of the 2-colex $L$, we associate a qudit with dimension $|G|$ and an orthonormal basis labelled by elements $g\in G$. Define operators corresponding to group multiplication and projection that act at a site as follows
		\begin{align}
			X_+^h\ket{g}&=\ket{hg}\\
			X_-^h\ket{g}&=\ket{gh^{-1}}\\
			T^h\ket{g}&=\delta_{h=g}\ket{g}
		\end{align}
		
		As compared to the qubit case, a general group requires a distinction between left and right multiplication, hence we have introduced both $X_+$ and $X_-$ to distinguish these two operations. Note that $[X_+^g,X_-^h]=0$, though $[X_+^g,X_+^h]\neq 0\neq [X_-^g,X_-^h]$ in general.

		A further group theoretic concept that will be useful is the commutator subgroup $[G,G] = \left<[g,h]:g,h\in G\right>$ where $[g,h] = g^{-1}h^{-1}gh$. This subgroup is normal, and the quotient group $G/[G,G]$ is the abelianization of $G$. A particularly useful property of $[G,G]$ is that it can alternatively be defined as the set of elements in $G$ which can be written as some product $g_1g_2g_3\cdots g_n$ that may be reordered such that it evaluates to identity.

\begin{figure}
		\centering
		\subfloat{
		\includegraphics{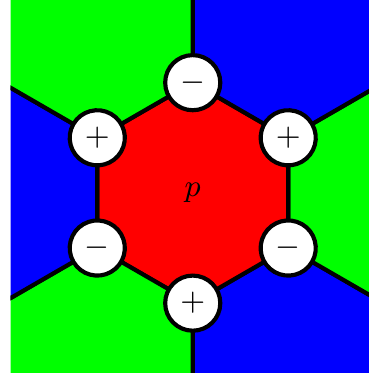}
		}
		\hspace{1cm}
		\subfloat{
		\includegraphics{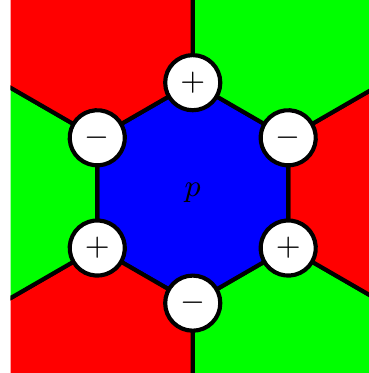}
		}
		\hspace{1cm}
		\subfloat{
		\includegraphics{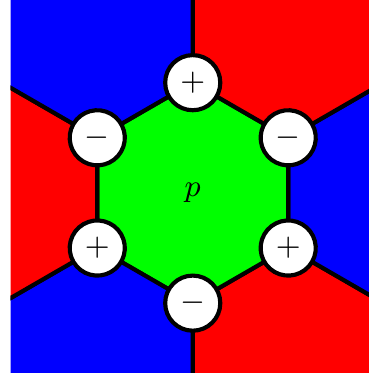}
		}
		\caption{Sign conventions for $X$-type stabilizers for example red, green, and blue plaquettes. These stabilizers are constructed from products of $X$ operators at each site around the plaquette. Those vertices with a $+$ denote left multiplication ($X_+$), while those with a $-$ denote right multiplication ($X_-$). These signs are determined by the order of plaquette colors around a site.}
		\label{f:stabconvention}
		\end{figure}

		In analogy to the qubit color code, the $G$-color codes are specified by a number of stabilizer operators. As in the qubit case, each plaquette will have an associated $X$- and $Z$-type stabilizer. However, in general, the form of these stabilizers will depend on the color of the plaquette under consideration. With this in mind, we can define $X$-type stabilizers for a plaquette as follows
		\begin{align}
			S^X_p  &= \frac{1}{|G|}\sum_{g\in G} A_p^g\\
			A_p^g &= \prod_{i\in p} X^g(i)\label{e:rightleftchoice}
		\end{align}
		where in \Eref{e:rightleftchoice} the $X^g$ operator acts either as left ($X_+^g$) or right ($X_-^g$) multiplication depending on the relative orientation of the three colors at a given site. Explicitly, when considering the $X$-type stabilizer corresponding to a blue or green plaquette, if the colors of plaquettes around a vertex are ordered $\{r,g,b\}$ when traversed clockwise, then the operator appearing at that vertex will be of the form $X_+$. Similarly, if the plaquette colors around the vertex are ordered $\{r,b,g\}$ when traversed clockwise, then the operator appearing at that vertex will take the form $X_-$. When considering red plaquettes, these conventions are reversed, so that $\{r,g,b\}$ clockwise ordering corresponds to $X_-$ and $\{r,b,g\}$ corresponds to $X_+$. Concrete examples of these conventions are illustrated in \Fref{f:stabconvention}.
		
		These conventions are chosen so that the $S^X$ operators corresponding to plaquettes of the privileged color (red) commute with the $S^X$ for both blue and green plaquettes. Making all three colored $X$-type stabilizers commute pairwise is impossible, and so the blue and green $S^X$ will not generally commute with each other for neighbouring plaquettes. Although this may seem at first glance to be a severe problem, it so happens that there exists a common $+1$ eigenspace of all $S^X$ regardless.
		
%------------------------------------------------------------------------------------------------------------%
		
%------------------------------------------------------------------------------------------------------------%
				
		The $Z$-type stabilizers for each of the three plaquette color are then defined as
		\begin{align}
			S^Z_{p,\mathrm{red}}&=\sum_{\prod g_i\in [G,G]}\prod_{i\in p} T^{g_i}(i)\label{e:szdefr}\\
			S^Z_{p,\mathrm{blue}}&=\sum_{\prod_{\mathrm{ACW}} g_i=e}\prod_{i\in p} T^{g_i}(i)\\
			S^Z_{p,\mathrm{green}}&=\sum_{\prod_{\mathrm{CW}} g_i=e}\prod_{i\in p} T^{g_i}(i)\label{e:szdefg}
		\end{align}
		where the ACW or CW denotes the product being taken anti-clockwise or clockwise around the plaquette, respectively. The origin of the product can easily be seen not to affect these operators. Additionally, the order of multiplication in $S^Z_{\mathrm{red}}$ does not affect the outcome. Since $[G,G]$ can be defined as the set of elements that can be decomposed into a product $h_1h_2h_3\cdots h_n$ such that it can be rearranged to evaluate to identity, whether or not some product of elements $\prod g_i$ is in $[G,G]$ will be independent of the order of multiplication. 
		
		When considering these $Z$-type stabilizers, it is clear the sense in which the red plaquettes are privileged. Their structure is not related to the full group $G$. Rather, we can consider $[G,G]$ as the preimage of the identity element of $G/[G,G]$ under the quotient map. Thus the structure of the red $Z$-type stabilizer is derived from that of the abelianization of $G$, as opposed to $G$ itself. For this reason, we will sometimes refer to red plaquettes as ``abelianized''. It should also be clear why we identified blue and green plaquettes as ``anticlockwise'' and ``clockwise'' respectively.
		
		The reason that we have defined the orderings of products in $S^Z$ operators as above is that this allows these operators to commute with all the $S^X$, as can easily be verified. That we are unable to preserve this commutativity without abelianizing the red plaquettes can be seen as a consequence of the fact that the blue and green $S^X$ fail to commute. It is impossible to preserve the full group structure of the red plaquette and have its $S^Z$ commute with both the blue and green $S^X$. In order to avoid this non-commutativity, we have reverted to the simpler structure of the abelianization of $G$, where no such problems exist, for the red plaquettes.
		
		In abelianizing the red plaquette $Z$-stabilizers, we have introduced an additional extensive degeneracy to the system. This degeneracy is local in the sense that it can be lifted through the addition of an extra class of local stabilizer operator that has no counterpart in the qubit case (or in fact for any abelian $G$-color code), where $[G,G]=\{e\}$. We call these $C$-type stabilizers, and they are defined for each red link as
		\begin{align}
			S_{l,\mathrm{red}}^C  &=\frac{1}{|[G,G]|}\sum_{n\in [G,G]}C_{l,\mathrm{red}}(n)\\
			C_{l,\mathrm{red}}(n) &= X_+^n(l,\uparrow)\otimes X_-^n(l,\downarrow)
		\end{align}
		where $X_{\pm}^n(l,\uparrow\!\downarrow)$ acts on the upper ($\uparrow$) or lower ($\downarrow$) qudit of the red link $l$ after it has been oriented such that the blue plaquette (ACW) is on the left of the link and the green plaquette (CW) on the right. It can be seen that the $S^C$ will commute with all $S^Z$ and $S^X$. By requiring the ground space to be in the $+1$ eigenspace of the $S^C$ as well as the $S^Z$ and $S^X$, the local degeneracies caused by abelianization of the red plaquettes are lifted.
		
		As noted, the $S^X$, $S^Z$, and $S^C$ stabilizers as written do not commute (as the $S^X$ do not in general commute). They nonetheless can be cast in the monomial stabilizer framework~\cite{VandenNest2011}, and have an associated frustration free ground space. However, commutativity can be restored by restricting the $S^X$ stabilizers for green and blue plaquettes to the $+1$ eigenspace of the $S^C$. These modified stabilizers take the form
		\begin{align}
			\tilde{S}^X_{p,\mathrm{green}}= S_p^X\prod_{l\in p\cap\mathrm{red}}S^C_l\\
			\tilde{S}^X_{p,\mathrm{blue}}= S_p^X\prod_{l\in p\cap\mathrm{red}}S^C_l
		\end{align}
		where $p\cap\mathrm{red}$ denotes all red links bounding the plaquette $p$. This will then give a set of commutative stabilizer operators generated by the $S^Z$, $\tilde{S}^X$ and $S^C$ operators, allowing us to write the Hamiltonian of a $G$-color code as
		\begin{align}
			H = -\sum_p\left(\tilde{S}^X_p +S^Z_p\right)-\sum_{\mathrm{red}\;l}S^C_l
		\end{align}	
		
			The ground space will be the common $+1$ eigenspace of all the stabilizer operators, and will be protected from excited states by a constant gap.		
		
		This completes the definition of the $G$-color code for the special case that the parity function is set to $s_i=1$ at every site $i$ of the lattice. The parity at a given site may be reversed by making the unitary transformation $\ket{g}\to\ket{g^{-1}}$ at that site. This completely exhausts the freedom we have in choosing the $S^X$, $S^Z$ and $S^C$ consistently. When viewed in this way, we can see the parity function as being analogous to the edge direction in the quantum double models, where reversing an edge is equivalent to applying the unitary taking $g\to g^{-1}$ on that edge. We will henceforth restrict to the $s_i=1$ parity case, with the understanding that all results will hold for any possible alternative parity choice.

%------------------------------------------------------------------------------------------------------------%

\section{Equivalence to copies of the quantum double models}\label{s:qdmap}
		
		A property of the qubit color code that will be particularly useful to us in understanding the $G$-color code is that it is locally equivalent to 2 copies of the toric code~\cite{Bombin2011,Bombin2011a,Haah2013}. We will now sketch an alternative proof of this fact (for a particular 2-colex)\footnote{The alternative mapping we describe is implicitly defined in \cite{Brell2011}.} and demonstrate how it generalizes to a general $G$-color code.
		
		\subsection{Qubit color code $\leftrightarrow$ toric code equivalence}

			Before we present the mapping between the color code and the toric code, we first briefly review the toric code~\cite{KitaevTC97}. For our purposes, it is sufficient to define the toric code on a square lattice, with qubits on edges. As in the color code, this model is defined by a number of stabilizer operators. Explicitly, for each vertex $v$ and plaquette $p$ of the lattice, we have
			\begin{align}
				K^X_v = \frac{1}{2}\left(1+\prod_{i\sim v}X_v\right)\\
				K^Z_p = \frac{1}{2}\left(1+\prod_{i\sim p}Z_p\right)
			\end{align}
			where we use notation such that $i\sim v$ runs over all qubits incident to $v$, and $i\sim p$ runs over all qubits bordering $p$. The toric code Hamiltonian is then given by
			\begin{align}
				H_{\mathrm{toric}} = -\sum_vK^X_v-\sum_pK^Z_p
			\end{align}
			in a very similar fashion to the color code.

\begin{figure}
		\centering
		\subfloat{
		\includegraphics{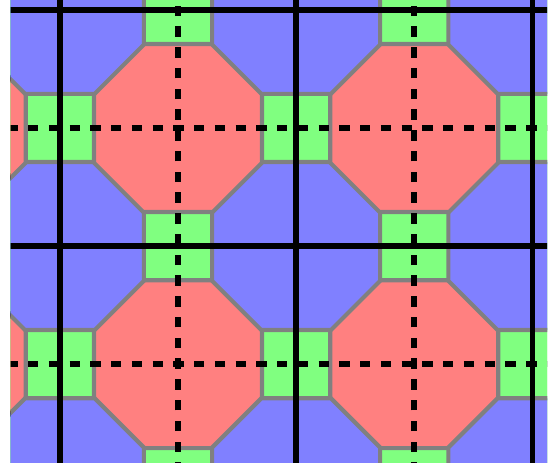}
		}
		\caption{Overlaid on the 4.8.8 color code lattice, the solid lines correspond to the edges of the first toric code lattice, while the dashed lines constitute the second toric code lattice.}
		\label{f:toriccolorcorrespond}
		\end{figure}

			For simplicity, we present a mapping between the color code and two copies of the toric code on a particular 2-colex. Specifically, we consider the 4.8.8 lattice, where the square plaquettes are colored green and the octagonal plaquettes blue and red in such a way as to satisfy 3-colorability (\Fref{f:488lattice}). We will show how to interpret a qubit color code on this 2-colex as 2 copies of the toric code. That is, we will find a local unitary map that transforms color code stabilizers to toric code stabilizers or actions on uncoupled ancilla systems. Intuitively, the mapping we present treats each green plaquette as 2 encoded qubits, one belonging to each copy of the toric code. The blue plaquettes will correspond to vertices (plaquettes) of the first (second) copy of the toric code, while the red plaquettes will correspond to plaquettes (vertices) of the first (second) copy of the toric code (see \Fref{f:toriccolorcorrespond}).

%------------------------------------------------------------------------------------------------------------%
%------------------------------------------------------------------------------------------------------------%			
			
			Consider the 4 qubits belonging to each green plaquettes as a stabilizer code with a 4-fold degenerate codespace. The stabilizers of this code are simply the green face stabilizers of the color code, i.e.
		\begin{align}
			S^X_{\mathrm{green}} &= \frac{1}{2}\left(1+\siteN{X}{X}{X}{X}\right)\\
			S^Z_{\mathrm{green}} &= \frac{1}{2}\left(1+\siteN{Z}{Z}{Z}{Z}\right)
		\end{align}
		where we use a graphical notation for the 4 operators acting on the vertices of the green plaquette, so that $\siteN{A}{B}{C}{D} \equiv A\otimes B\otimes C\otimes D$ on the 4 relevant qubits.
	
			This 4-dimensional codespace corresponding to each green plaquette can be considered as 2 encoded qubits, one for each copy of the toric code. In this way, the qubits of the new toric code lattices have a direct correspondence to the green plaquettes of the color code lattice. It will now be convenient to bipartition these green plaquettes into those with red plaquettes on their right and left ($h$-type) and those with red plaquettes above and below them ($v$-type).
			
			On $h$-type green plaquettes, encoded Pauli algebras can be defined for each of the 2 encoded qubits as
			\begin{align}
				X_{\mathrm{enc}}^h(1) &= \siteN{X}{X}{I}{I}\\
				X_{\mathrm{enc}}^h(2) &= \siteN{X}{I}{I}{X}\\
				Z_{\mathrm{enc}}^h(1) &= \siteN{Z}{I}{I}{Z}\\
				Z_{\mathrm{enc}}^h(2) &= \siteN{Z}{Z}{I}{I}
			\end{align}
			where $X_{\mathrm{enc}}^h(2)$ acts as Pauli $X$ on the $2\nd$ encoded qubit for the $h$-type green plaquette under consideration.
			
			For $v$-type green plaquettes, the encoded operators can be defined by 
			\begin{align}
				X_{\mathrm{enc}}^v(1) &= X_{\mathrm{enc}}^h(2)\\
				X_{\mathrm{enc}}^v(2) &= X_{\mathrm{enc}}^h(1)\\
				Z_{\mathrm{enc}}^v(1) &= Z_{\mathrm{enc}}^h(2)\\
				Z_{\mathrm{enc}}^v(2) &= Z_{\mathrm{enc}}^h(1)
			\end{align}
			That is, on $v$-type plaquettes, the definitions of the first and second encoded qubits are exchanged.
			
			Note that these definitions are not unique. In particular, because the codespace is the $+1$ eigenspace of $S^X_{\mathrm{green}}$ and $S^Z_{\mathrm{green}}$, multiplying any operator by $\siteN{X}{X}{X}{X}$ or $\siteN{Z}{Z}{Z}{Z}$ is a trivial operation, and so the encoded operators are invariant under $180^\circ$ rotations.

			One could more rigorously define the unitary enacting this encoding as a function from operators on the 4 qubits of each green plaquette to the 2 encoded (toric code) qubits, as well as 2 ancilla qubits. Notably, this transformation would take green color code stabilizers to an action on the ancilla space, and so they become ``trivial'' in the sense of Ref.~\cite{Bombin2011a}.
			
			In terms of the encoded toric code qubits, the action of the color code stabilizers for blue and red plaquettes is
			\begin{align}
				S^X_{\mathrm{red}} \to K^X(1)\\
				S^Z_{\mathrm{red}} \to K^Z(2)\\
				S^X_{\mathrm{blue}} \to K^X(2)\\
				S^Z_{\mathrm{blue}} \to K^Z(1)
			\end{align}			
			if we interpret the lattices of the two toric codes as in \Fref{f:toriccolorcorrespond}. Thus the unitary we have described maps all the stabilizers of the original color code model to stabilizers of the toric code model, or operators on uncoupled ancilla qubits. This completes the demonstration of equivalence.

		\subsection{$G$-color code $\leftrightarrow$ quantum double model equivalence}\label{s:qdcccorrespond}

			The equivalence between the qubit color code and the toric code  on the 4.8.8 lattice shown above can be generalized to an equivalence between the $G$-color code and 2 quantum double models. However, as compared to the qubit case (which corresponds to $G=\Z_2$), the two quantum double models will in general be different. One will be based on the group $G$, while the other will be based on the abelianization of the group $G/[G,G]$.

\begin{figure}
			\centering
			\subfloat{
			\includegraphics{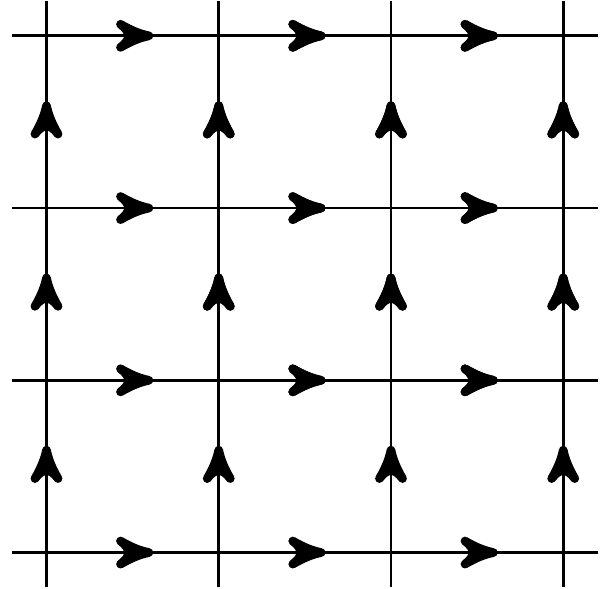}
			}
			\caption{A directed square lattice, consistent with the quantum double stabilizers defined in Eqns. (\ref{e:kvxqd}) - (\ref{e:kpzqd}).}\label{f:dirsquarelatt}
			\end{figure}

			Before we continue, we briefly review the quantum double models~\cite{KitaevTC97}. Conventionally, these models are defined on a directed lattice, with edge direction playing a similar role to our parity function. We will define a quantum double model for a group $G$ on an particular directed square lattice (\Fref{f:dirsquarelatt}), with qudits of dimension $|G|$ placed at every edge, noting that transforming the qudit on an edge by $g\to g^{-1}$ is equivalent to reversing the direction of the edge (qudits have a natural basis labelled by $g\in G$). 
			
%------------------------------------------------------------------------------------------------------------%
%------------------------------------------------------------------------------------------------------------%	
			
			At each vertex $v$ and plaquette $p$ of the lattice, the stabilizers of the quantum double model can be written as
			\begin{align}
				K^X_v &= \sum_g A_v(g)\label{e:kvxqd}\\
				A_v(g) &= X^g_{-,U} X^g_{-,R} X^g_{+,D} X^g_{+,L}\label{e:avqd}\\
				K^Z_p &= \sum_{g_1g_2g_3g_4=e}\prod_{i\in p} T^{g_1}_U T^{g_2^{-1}}_R T^{g_3^{-1}}_D T^{g_4}_L\label{e:kpzqd}
			\end{align}
			where the subscript of the $X$ operators in (\ref{e:avqd}) and the $T$ operators in (\ref{e:kpzqd}) denote whether they act on the qubit $U$p, $R$ight, $D$own, or $L$eft from the center of the plaquette or vertex under consideration.			
			
			As was the case for the qubit color code $\leftrightarrow$ toric code mapping (the $\Z_2$ case), we will explicitly consider only the 4.8.8 lattice, where the square plaquettes are colored green, and the octagonal plaquettes are colored red and blue (\Fref{f:488lattice}). The intuition for our construction is much the same, in that we will encode the quantum double degrees of freedom in the qudits at the vertices of the green plaquette. However, particularly in the case of non-abelian group $G$, the mapping will be less straightforward.
			
		As before, green plaquettes are labelled $h$- or $v$-type depending on whether red plaquettes are at their sides or above and below them. Given this, define the codespace for $h$-type green plaquettes via the stabilizers of the plaquette
		\begin{align}
			S^{Z,h}_{\mathrm{green}} &= \sum_{g_1g_2g_3g_4=e}\siteN{T^{g_1}}{T^{g_2}}{T^{g_3}}{T^{g_4}}\\
			S^{C,h}_{l_1,\mathrm{green}} &=\sum_{n\in[G,G]}C^h_{l_1,\mathrm{green}}(n)\\
			S^{C,h}_{l_2,\mathrm{green}} &=\sum_{n\in[G,G]}C^h_{l_2,\mathrm{green}}(n)\\
			\tilde{S}^{X,h}_{\mathrm{green}}&=S^{C,h}_{l_1,\mathrm{green}}S^{C,h}_{l_2,\mathrm{green}}\sum_gA^h_{\mathrm{green}}(g)
		\end{align}
		with
		\begin{align}
			A^h_{\mathrm{green}}(g) &=\siteN{X_+^g}{X_-^g}{X_+^g}{X_-^g}\\
			C^h_{l_1,\mathrm{green}}(n)&= \siteN{X_+^n}{I}{I}{X_-^n}\\
			C^h_{l_2,\mathrm{green}}(n)&= \siteN{I}{X_-^n}{X_+^n}{I}
		\end{align}
		
		On $v$-type green plaquettes, the stabilizers can be found by rotating the $h$-type stabilizers by $90^\circ$. These definitions can be seen to be invariant under $180^\circ$ rotations.
		
		Given this codespace at each $h$-type green plaquette, we can write encoded operators on this space as
		\begin{align}
			X^g_+(1)\equiv& \siteN{X^g_-}{X^g_+}{I}{I} &
			T^g(1)\equiv& \sum_{g_2g_3 = g}\siteN{I}{T^{g_2}}{T^{g_3}}{I} \\
			X^g_-(2) \equiv& \sum_{k\in g[G,G]}\siteN{I}{X^k_-}{X^k_+}{I} &
			T^g(2) \equiv& \sum_{g_1g_2 \in g[G,G]}\siteN{T^{g_1}}{T^{g_2}}{I}{I} \label{e:2encqudit1}\\
			X^g_-(1)\equiv& \siteN{I}{I}{X^g_-}{X^g_+} &
			T^g(1) \equiv& \sum_{g_4g_1 = g^{-1}}\siteN{T^{g_1}}{I}{I}{T^{g_4}}  \\
			X^g_+(2) \equiv& \sum_{k\in g[G,G]}\siteN{X^k_+}{I}{I}{X^k_-} &
			T^g(2) \equiv& \sum_{g_3g_4 \in g^{-1}[G,G]}\siteN{I}{I}{T^{g_3}}{T^{g_4}} \label{e:2encqudit2}
		\end{align}
		
		As in the $\Z_2$ case, we can define logical operators for the encoded qudits in multiple equivalent ways within the codespace. We have written two particularly useful sets of logical operators for each of the encoded qudits here. These encoded operators can be seen to commute with the stabilizers defined above. The encoded operators acting on the $v$-type green plaquettes can again be found by rotating the $h$-type operators $90^\circ$ clockwise.
		
		In the $\Z_2$ case, the two encoded systems (labelled $1$ and $2$) were both of the same dimension (they were both encoded qubits). In general, however, this is not the case. Examining the encoded operators on qudit 2 (\ref{e:2encqudit1}) or (\ref{e:2encqudit2}), we can see that states in this space are labelled only by $k\in G/[G,G]$, and thus this qudit is $|G/[G,G]|$ dimensional. This can be seen by noting that $g[G,G]$ is the coset of $G$ corresponding to a particular element of $G/[G,G]$, and in particular $g[G,G]=ng[G,G]$ for any $n\in [G,G]$. In contrast, qudit 1 can be seen to have basis labelled by $g\in G$, and is thus $|G|$ dimensional.

\begin{figure}
		\centering
		\subfloat{
		\includegraphics{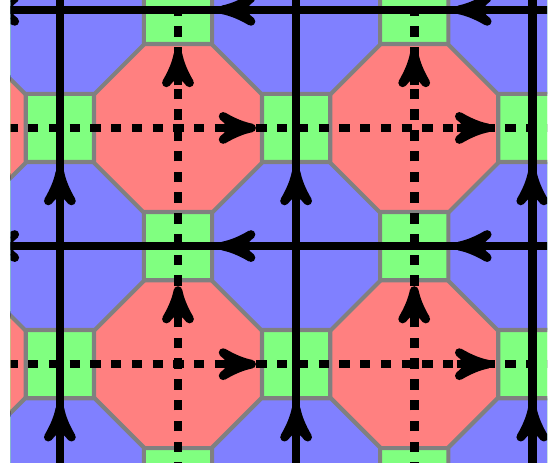}
		}
		\caption{Overlaid on the 4.8.8 $G$-color code lattice, the solid lines correspond to the edges of the first quantum double lattice, while the dashed lines constitute the second quantum double lattice.}
		\label{f:qdcorrespond}
		\end{figure}

		In terms of this encoding, it is clear that once again, the $X$-type stabilizers of the $G$-color code corresponding to green plaquettes can be thought of as acting only on some ancilla space (they have no action on the encoded space by construction). Similarly, the $C$-type stabilizers corresponding to red links of the $G$-color code act only within a green plaquette, and commute with the encoded operators, so they can also be interpreted as acting on an ancilla space. In contrast, the $G$-color code stabilizers for blue and red plaquettes can be rewritten in terms of the encoded qudits. Their action under the encoding map can be written as
			\begin{align}
				S^X_{\mathrm{red}} \to K^X(1)\\
				S^Z_{\mathrm{red}} \to K'^Z(2)\\
				\tilde{S}^X_{\mathrm{blue}} \to K'^X(2)\\
				S^Z_{\mathrm{blue}} \to K^Z(1)
			\end{align}		
			where $K'$ are defined by Eqs. (\ref{e:kvxqd}) and (\ref{e:kpzqd}) over the group $G/[G,G]$, as opposed to the $K$ which are defined over $G$. The $K'(2)$ and $K(1)$ operators act on the quantum double lattices depicted in \Fref{f:qdcorrespond}.
			
%------------------------------------------------------------------------------------------------------------%
%------------------------------------------------------------------------------------------------------------%				

			Thus the unitary we have described maps all the stabilizers of the $G$-color code model to stabilizers of the quantum double model, or operators on uncoupled ancilla qubits, in analogy to the qubit color code $\leftrightarrow$ toric code mapping. Thus we have demonstrated an equivalence between a $G$-color code and a $G$ quantum double model together with a $G/[G,G]$ quantum double model (equivalently, a single $G\times G/[G,G]$ quantum double model).
		
%------------------------------------------------------------------------------------------------------------%	

\section{Properties of Generalized Color Codes}\label{s:properties}

	The qubit color codes have a number of properties that make them of interest to the topological quantum information community. Here we will take a brief survey of some of the most important properties of the $G$-color codes. Where relevant, we will discuss the connection to the properties of the qubit color codes. Note that some properties of these models for abelian groups $G$ are explored in some depth in \cite{Sarvepalli2010}.

		\subsection{Anyon spectrum}
			
			The correspondence between the $G$-color codes and the quantum double models developed in \Sref{s:qdmap} allows us to immediately import results from the study of the quantum double models and interpret them in the context of the $G$-color codes. 
			
			Ref.~\cite{Bombin2011} shows that so-called topological stabilizer groups are equivalent iff they have isomorphic topological charges. This result does not apply directly to the equivalence that we have established between quantum double models and $G$-color codes because they are not Pauli stabilizer models. However, even without the level of rigor available for Pauli stabilizer models, we can still use the intuition behind this theorem to draw a correspondence between the anyonic content of the quantum double models and the $G$-color codes.
					
			It is easy to see that all local operations on the qudits of the quantum double models in \Sref{s:qdcccorrespond} can be mapped to some local operations on the qudits of the $G$-color code (though the converse is not so straightforward due to the presence of the ancillae in the mapping). Similarly, anyonic charges of the quantum double models can be mapped to charges living on the red and blue plaquettes of the $G$-color code. This allows us to state that the $G$-color code is supporting the anyons of both the $G$ and $G/[G,G]$ quantum double models. Notably, this includes non-abelian anyons for non-abelian $G$.
			
			In general, since the anyons of a quantum double model for group $G$ are given by the irreducible representations of the Drinfeld double of $G$, $\mathrm{Irrep}(\mathcal{D}(G))$, we would expect from this line of reasoning that the anyonic content of a $G$-color code is given by $\mathrm{Irrep}(\mathcal{D}(G\times G/[G,G]))$.
			
			Anyonic charges of the quantum double models can be created and moved by ribbon operators~\cite{KitaevTC97}. Each of these ribbon operators must have a corresponding ribbon operator creating or moving the analogous charges on the $G$-color code. Given these operators, we could imagine braiding charges in the $G$-color code corresponding to any desired braiding in a quantum double model. In particular, we could perform braiding of a non-abelian anyons that implements universal quantum computation in $G$-color codes for $G$ non-nilpotent~\cite{Mochon2003,Mochon2004}.
				
%------------------------------------------------------------------------------------------------------------%		
		
		\subsection{Further implications of equivalence}

			A concrete motivation for demonstrating equivalence between two topological models is that it allows decoding or error correcting routines for one model to apply to a broader class of models~\cite{Bombin2011a}. In our case, this amounts to the observation that a $G$-color code can be decoded if an equivalent procedure for decoding the quantum double model for group $G$ is known. Although decoding for these non-abelian anyon models has yet to be explicitly demonstrated (though related work has been shown~\cite{Wootton2014, Brell2013}), by equating our $G$-color codes to the well established quantum double models, we can exploit any results in terms of decoding that are available for them.
			
			Ref.~\cite{Bombin2011} shows that all Pauli stabilizer codes are equivalent to copies of the toric code. The fact that we are able to demonstrate equivalence between quantum double models and our $G$-color codes suggests that a more general equivalence may hold whenever the models are constructed from commuting projectors based on the $X$ and $T$ operator algebra for a group $G$. This may also point to a useful restriction of the monomial stabilizer formalism~\cite{VandenNest2011} when the desired algebra structure is related to a particular group.
			
			Note that we have technically not demonstrated equivalence between an arbitrary $G$-color code and quantum double models, as our mapping is specifically tailored to the 4.8.8 lattice. This was sufficient to prove that the $G$-color codes are capable in principle of supporting quantum double anyons, but we have not shown that this is true for a general lattice. However, the general principles of topologically ordered systems suggest that the microscopic lattice details should not affect global properties of the system such as its anyon spectrum, and so we feel confident in taking this correspondence to hold in general.
		
%------------------------------------------------------------------------------------------------------------%		

	\subsection{Degeneracy and Boundaries}\label{s:boundary}

		As is generic for topological codes, the degeneracy of a particular code is highly dependent on the topology of the surface in which it is embedded, or, in the planar case, the particular choice of boundary conditions. On closed (orientable) manifolds, this degeneracy is independent of the microscopic details and can be derived from the anyon spectrum~\cite{Nayak2008}. This means that the degeneracy of the $G$-color codes on closed manifold can directly be calculated as the degeneracy of the quantum double model for $G\times G/[G,G]$.
		
		Similarly, characterization of the possible gapped boundaries for topologically ordered models is based on the anyon spectrum rather than microscopic details. Possible boundary types and their properties can thus be found by appealing to the known results for the quantum double models~\cite{Beigi2011,Kitaev2011,Hung2014}. Since this theory is already well established in general, there is no reason to delve into it here. However, before moving on we will briefly discuss some special boundary types that are natural in the context of the color codes, and planar codes that can be constructed from them.
		
\begin{figure}
		\centering
		\subfloat[]{\label{f:cctribound}
		\includegraphics{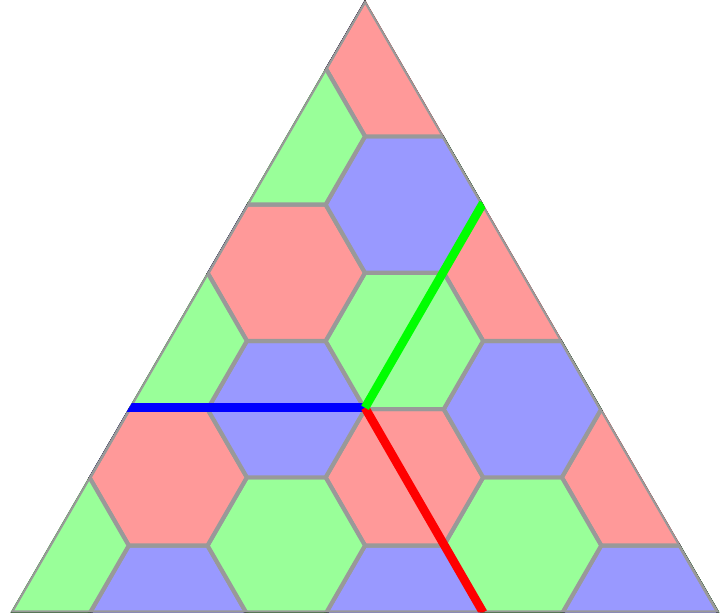}
		}
		\hspace{-1cm}
		\subfloat[]{\label{f:ccrectbound}
		\includegraphics{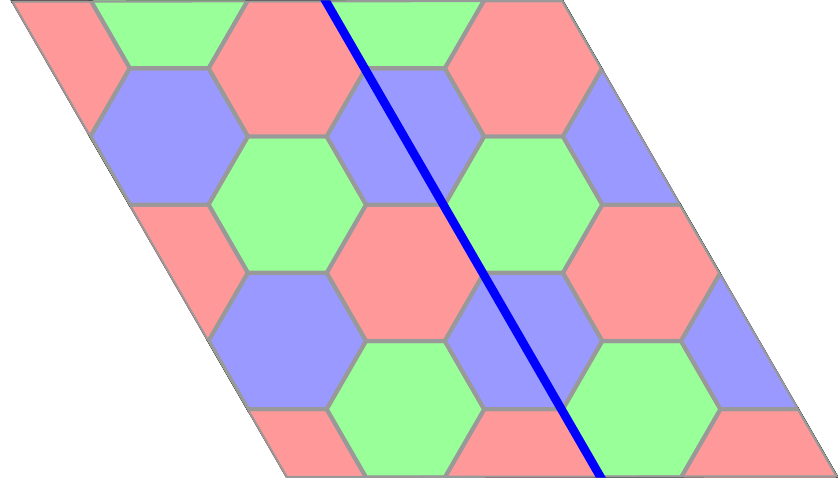}
		}
		\caption{Two examples of planar code boundaries, together with examples of logical operator strings on these lattices. (a) A triangular boundary. (b) A rectangular boundary.}
		\label{f:ccbounds}
		\end{figure}

		The most common planar qubit color codes are triangular~\cite{Bombin2006} (\Fref{f:cctribound}). Each side of the triangle is associated with the color not appearing in plaquettes on that boundary. These triangular color codes encode one qubit. Small triangular color codes have been experimentally prepared and manipulated~\cite{Nigg2014}. Another common form of planar color code is rectangular, as in \Fref{f:ccrectbound}. In this case, two logical qubits are encoded. More generally, a natural boundary for the color code is labelled by a color according to the color not appearing on plaquettes at that boundary. Degeneracy is introduced to the system when the lattice boundary consists of several different colored segments. Although these boundaries arise naturally in the study of color codes, they are not the only boundary types that can occur. In particular, note that the number of boundary types we have just described is 3 (labelled by colors), while a topologically ordered model equivalent to two copies of the toric code can give rise to 6 types of boundary~\cite{Bravyi1998, Beigi2011, Kitaev2011, Barkeshli2013a, Barkeshli2013b, Levin2013}). The reason that the boundaries we described are particularly natural is that they require no special modification to the stabilizer operators (only the 2-colex on which the model is defined).
		
%------------------------------------------------------------------------------------------------------------%
		
%------------------------------------------------------------------------------------------------------------%	
		
		In the general case of a $G$-color code, the effect of boundaries is a little more complicated due to the asymmetry between the three colors (and of course the more complicated algebraic structure). As before, degeneracy is introduced to the codespace when more than two distinct boundaries exist, but the counting of this degeneracy depends on the color (or type) of the boundaries. Since the general case can be determined by appealing to results for the quantum double models, we will simply describe the logical operators and degeneracy of three distinct planar $G$-color codes: the blue-green rectangular codes (i.e.~one with alternating blue and green boundaries as shown in \Fref{f:ccrectbound}), the blue-red rectangular codes, and the triangular codes. Up to the equivalence between the blue-red rectangular code and a green-red rectangular code, this exhausts all possible rectangular and triangular codes.
		
		In order to describe the logical operators in these codes explicitly, it would be necessary to construct ribbon operators as in the quantum double models~\cite{KitaevTC97, Bombin2008a}, which is straightforward but tedious. Instead, we will simply state the types of logical operators that arise and their relationships in order to calculate the degeneracy of these codes.		
		
		\subsubsection{Blue-green rectangular code}
		
			Before discussing the logical operators of these codes, we should quickly note a subtlety in their definition. At each corner between a blue and green edge, there is a red plaquette with one qudit that touches neither green nor blue plaquettes. This qudit should be understood to have a $C$-type stabilizer associated with it, as would be the case if there were an edge emanating from this qudit.
		
			Between the two blue boundaries of this code, it is possible to construct a set of $X$-type operators labelled by group elements in $G$ that we call $X_{\mathrm{log,blue}}^g$. These would (largely) run along blue links of the lattice. However, note that along a green boundary, the distinction between blue links and red links disappears. Thus along this boundary, we can construct a family of operators from $C$-type stabilizers, each labelled by $n\in [G,G]$ that runs along blue links connecting the blue boundaries. After accounting for equivalence up to these operators, only blue $X$-type logical operators labelled by $k\in G/[G,G]$ are independent. Similarly we can construct a set of independent green logical $X$ operators labelled by $k\in G/[G,G]$, $X_{\mathrm{log,green}}^k$. These operators will commute pairwise.
			
			We can also construct $Z$- (or $T$)-type logical operators between the blue boundaries. However, if we were to construct an entire set $Z_{\mathrm{log,blue}}^g$ labelled by each $g\in G$, these would not commute with the stabilizers. Instead, we can only construct independent representatives labelled by each $k\in G/[G,G]$. Similar considerations apply to the $Z_{\mathrm{log,green}}^k$.
			
			We can then form two pairs of logical qudit algebras generated by $X_{\mathrm{log,green}}^k$ with $Z_{\mathrm{log,blue}}^k$ and $X_{\mathrm{log,blue}}^k$ with $Z_{\mathrm{log,green}}^k$ for $k\in G/[G,G]$. We thus say the logical qudits associated with each of these sets of operators are abelianized. The total degeneracy is $|G/[G,G]|^2$.
		
		\subsubsection{Blue-red rectangular code}		
		
		As in the blue-green rectangular code, we can define $X_{\mathrm{log,blue}}^g$. However, since there is no green boundary, these are now independent for all $g\in G$. In contrast, red $X$-type logical operators are only defined for each $k\in G/[G,G]$ (again due to equivalence under $C$-type stabilizers).
		
		Conversely, the red $Z$-type operators are defined for each $g\in G$, while the blue $Z$-type operators are only independently defined for each $k\in G/[G,G]$. This leads to logical qudit algebras generated by $X_{\mathrm{log,blue}}^g$ with $Z_{\mathrm{log,red}}^g$ for each $g\in G$ and $X_{\mathrm{log,red}}^k$ with $Z_{\mathrm{log,blue}}^k$ for each $k\in G/[G,G]$. In this case only one of the logical qudits has been abelianized, and the total degeneracy is $|G|\cdot|G/[G,G]|$.
		
		\subsubsection{Triangular codes}
		
		For the triangular code, it is not so straightforward to assign a color type to each logical operator because they may branch. Despite this, by constructing logical operators that run entirely along one side of the triangle, it can be seen that only independent $X_{\mathrm{log}}^k$ and $Z_{\mathrm{log}}^k$ for each $k\in G/[G,G]$ may be defined, and so determine the degeneracy of these codes to be $|G/[G,G]|$.

	\subsection{Topological defects}

		Topological defects can play an important role in topological systems. In particular, braiding of certain types of defects can implement quantum computation (through code deformation) in much the same way as braiding of anyons. Particular types of topological defects such as holes~\cite{Bombin2009a} and twists~\cite{Bombin2010a, Bombin2011b} have been developed for this purpose in simple models such as the toric codes or qubit color codes.	As with our discussion of the anyon spectrum of $G$-color codes, we can import many known results from quantum double models into our study of topological defects in these models, taking the equivalence between quantum doubles and $G$-color codes to be general (i.e.~independent of the details of the lattice).
		
		The theory of topological defects is closely related to the theory of boundaries. The study of domain walls (general boundaries) between two topological phases can be used to explore all the topological defects we will consider here. This theory is developed for the quantum double models in \cite{Beigi2011} (see also \cite{Kitaev2011} for related work). By making use of the correspondence between the $G$-color codes and the quantum double models, we can import the characterization of the domain walls possible between two phases of $G$-color codes.
		
		Twists are topological defects at which domain walls terminate~\cite{Kitaev2011}. They induce an automorphism of the set of anyons, such that an anyon braiding around a twist will return as a different species (as dictated by the automorphism). Twists have well-defined fusion and braiding amongst themselves, and can be used for topological quantum computation~\cite{Bombin2010a, Bombin2011b}. Possible twists for quantum double models (and hence $G$-color codes) are studied in Refs.~\cite{Beigi2011,Kitaev2011}.
				
		The second type of defects we consider are holes (or punctures). These are closed boundaries between the $G$-color code and vacuum (the topologically trivial phase) within a planar topological code. They can also have well-defined fusion and braiding relations, and can be used for topological quantum computation~\cite{Bombin2009a}. Again, the characterization of such hole types can be imported from the boundary theory of quantum double models, as discussed in \Sref{s:boundary}.
			
		Finally, condensation is a mechanism via which the topological order in an anyonic model can be changed. In a condensed phase, the anyon density for a particular species is given a non-trivial value in the ground space, and this can lead to interesting effects, e.g.~confinement of other anyon species. Condensation amounts to a deformation in the bulk of the model under consideration and again an analogy can be drawn to equivalent processes in quantum double models as studied in \cite{Bombin2008a} (see also \cite{Eliens2010, Kitaev2011, Kong2013}). These kinds of effects may also be used to perform quantum computation by manipulating the regions of condensed phases~\cite{Bombin2011nest}.
	
%------------------------------------------------------------------------------------------------------------%	

	\subsection{Transversality properties}
	
		One feature of the qubit color codes that is particularly appealing is its large set of transversal gates~\cite{Bombin2006} (in comparison to the toric code, for example). In fact, this feature is inherited for the color codes based on abelian groups~\cite{Sarvepalli2010}. The most important transversal gate that these abelian color codes possess is the Hadamard gate. However, as discussed in~\cite{Brellcluster}, operator algebras based on finite non-abelian groups will not have a counterpart to the Hadamard gate, due to the inequivalence of the group algebra and the corresponding representation algebra.
		
		In a similar vein, logical operators for codes with similar algebraic structure to the $G$-color codes (significantly the quantum double models) do not generally have large transversal gate sets. The transversal ``string''-like logical operators that appear for abelian groups generally become non-transversal ``ribbon'' operators in this context~\cite{KitaevTC97, Bombin2008a} when the topological charges of the model become non-abelian (and thus able to perform quantum computation by braiding). For these reasons, we do not expect the $G$-color codes to have particularly interesting transversal gate sets.

	\subsection{Construction from cluster state}
	
		The qubit color code can be constructed in a straightforward way from a suitable cluster state~\cite{Bombin2008}. We can view this relationship (appropriately generalized) as one way to define the $G$-cluster states.
				
%------------------------------------------------------------------------------------------------------------%
		\begin{figure}
		\centering
		\includegraphics{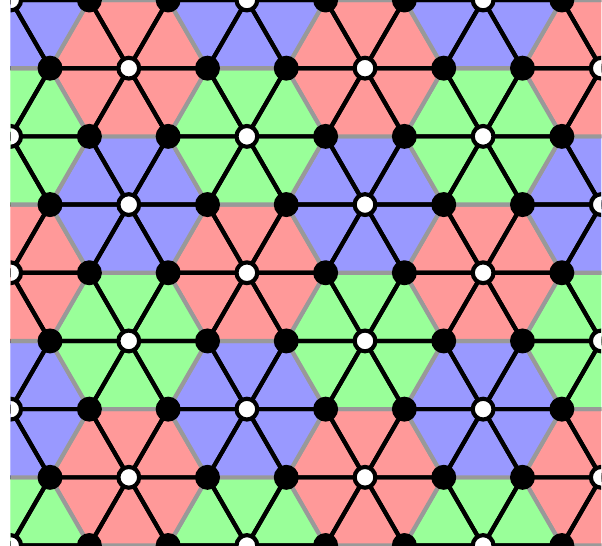}
		\caption{By preparing a generalized cluster state on the lattice shown with black edges, and projecting a subset of the qudits (those represented by open circles) into particular states, we can prepare a $G$-color code.}
		\label{f:clusterhex}
		\end{figure}
		
%------------------------------------------------------------------------------------------------------------%
	
	The qubit color code can be produced by beginning with a cluster state on the lattice shown in \Fref{f:clusterhex}, and projecting a subset of the qubits (corresponding to each plaquette of the color code) into the $\ket{0}$ state~\cite{Bombin2008}. This procedure can directly be generalized to produce $G$-color codes from generalized cluster states~\cite{Brellcluster} based on the group $G$. Given a suitably prepared generalized cluster state, the analogue of the $\ket{0}$ state projections would naively be expected to be a projection to $\ket{e}$. However, performing these projections would not give rise to the $G$-cluster states we have defined (the resulting states would not have local stabilizers).
	
	Instead, the projection on the qudits corresponding to red plaquettes must be generalized to projections to $\sum_{n\in[G,G]}\ket{n}$, while those on blue and green plaquettes would remain as projections to $\ket{e}$. This will result in the desired $G$-color code state. Of course for any abelian group $G$, $[G,G]=\{e\}$, and so we recover the standard qubit procedure for $G=\Z_2$.
		
	As discussed in Ref.~\cite{Brellcluster} for the similar case of the toric code, if we were to physically use measurements of a suitable basis in lieu of projections to the $\ket{e}$ or $\sum_{n\in[G,G]}\ket{n}$ states in an attempt to prepare $G$-color code states from generalized cluster states in the laboratory, the resulting state could be interpreted as an excited state of the $G$-color code with excitations determined by measurement outcomes.

\section{Discussion}\label{s:discussion}

	We have defined a generalization of the color codes to finite group $G$, and explored many of their basic properties. Some of the useful features of the qubit color codes do not carry over to the general case, most notably the large set of transversal gates. Nonetheless, we do not have many exactly-solvable models of topologically ordered systems and so this new family may be of interest as a testbed for topological phenomena, or may have properties that are difficult to find in existing models. Furthermore, the relationship between the color codes and many other interesting systems may allow for further extension of this work. We briefly discuss a few of the most obvious extensions below.			
	
	\subsection{Extensions of the model}

		\subsubsection{Topological Subsystem Codes}
		
\begin{figure}
			\centering
			\subfloat{
			\includegraphics{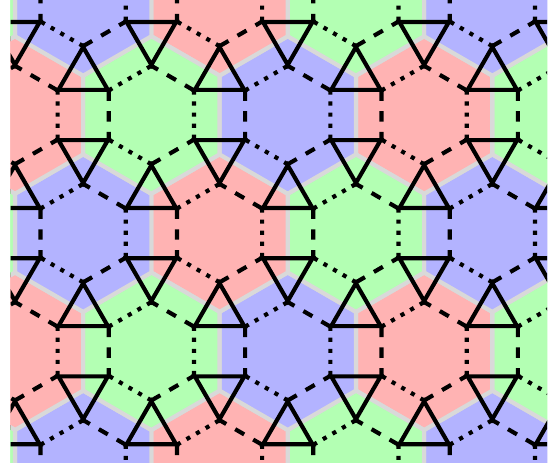}
			}
			\caption{The topological subsystem code is defined on an an expanded 2-colex lattice. In this example, we have expanded the 6.6.6 honeycomb lattice. The links of the expanded lattice come in three types, shown as dashed, dotted, and solid, corresponding to the three possible interaction types: $X$, $Y$, or $Z$.}
			\label{f:tssclattice}
			\end{figure}

			Topological subsystem codes are a family of models that are related to color codes, but whose Hamiltonians require only 2-body interactions~\cite{Bombin2009, Bombin2010}. They are defined on 2-colexes whose links and sites have been expanded to create a new lattice, as in \Fref{f:tssclattice}. Each edge of the expanded lattice carries an operator of the form $X\otimes X$, $Y\otimes Y$, or $Z\otimes Z$.
			
%------------------------------------------------------------------------------------------------------------%
		
%------------------------------------------------------------------------------------------------------------%			
			
			We will not go into details about these models, but we would not expect them to be generalizable to an arbitrary finite group $G$ as  we have done for color codes for the following reason. In our generalization, we have treated the $X$ operator as if it were a group multiplication operator. We could also have generalized the $Z$ operator as $Z^\Gamma_{ij} = \sum_g[\Gamma(g)]_{ij}T_g$, for $[\Gamma(\cdot)]_{ij}$ the $(i,j)\th$ matrix element of a representation $\Gamma$. In particular, for the group $\Z_2$ this gives $Z^\mathrm{triv}=I$ and $Z^\mathrm{alt}=\sigma_z$ for the trivial and alternating representations, respectively. In this way, we can interpret the $Z$ operators as acting like representations (see Ref.~\cite{Brellcluster} for a more detailed discussion).
			
			Given these interpretations of $X$ and $Z$, we can attempt to generalize a given CSS model. However, when presented with a non-CSS model such as the topological subsystem codes, the immediate problem of how to interpret a $Y$ operator in terms of group structure has no obvious answer. For the $\Z_2$ case (and indeed any cyclic group), there is a natural correspondence between the group elements and the representations of the group. This allows us to unambiguously define products of $X^gZ^g$ that can serve as a generalization of $Y$ (up to constants). For this reason, we would not expect significant obstacles to generalizing the topological subsystem codes to abelian groups. In contrast, for non-abelian groups we cannot generally rely on a natural correspondence between group elements and representations, and for this reason it seems unlikely that our strategy will provide a sensible generalization of the topological subsystem codes for these groups. We refer the interested reader to Ref.~\cite{Brellcluster} for a more comprehensive discussion of similar issues.

		\subsubsection{Higher dimensional models and gauge color codes}
		
			The 3D qubit color codes are introduced in \cite{Bombin2007}, and $D$-dimensional qubit color codes can also be defined~\cite{Bombin2009b} and generalized to gauge color codes~\cite{Bombin2013}. They are based on the notion of a $D$-colex in analogy to the 2-colex of the 2D model. We anticipate that the same algebraic structure used here to define the 2-dimensional $G$-color codes (such as the commutator subgroup) may be used to define generalizations of the higher-dimensional color codes in some cases.
			
			The gauge color codes are the most general setting for these higher-dimensional color codes, and are defined for spatial dimension $D$ by a $D$-colex and two positive integers $d,e$ such that $d+e\leq D$. These two integers specify the geometry of the generators of the stabilizer group (or more generally the gauge group in the language of subsystem codes~\cite{Kribs2005,Kribs2006}). In particular, the $Z$-type interactions in the Hamiltonian are associated with $(d+1)$-dimensional objects, while the $X$-type interactions are associated with $(e+1)$-dimensional objects. When generalizing these models to arbitrary finite group $G$ in the most naive way, it seems necessary to restrict to $d=1$, since the definition of the $S^Z$ stabilizers (\ref{e:szdefr}-\ref{e:szdefg}) requires these operators to be associated to an object with a notion of cycles, i.e.~a 2-dimensional face. The same considerations restrict the kinds of higher-dimensional generalizations that are possible for e.g.~the quantum double models and string net models, though more complex higher dimensional analogues can be defined~\cite{Walker2012}. However, the fact that the gauge color codes allow more general structures than the quantum double models (the analogous construction for quantum double models would not allow $d+e<D$) suggests that there may be novel ways to implement non-abelian topological orders in higher dimension with these methods.

		\subsubsection{Extension to more general algebras}
	
			Our model is very similar in construction to the quantum double models, as can be seen by the mapping of \Sref{s:qdmap}, which makes this correspondence explicit. Given that more general quantum double models can be defined based on Hopf algebras (and potentially more general algebraic structures)~\cite{Buerschaper2010,Buerschaper2010a}, a natural question arises whether these models, too, have color code counterparts. We expect that this may be possible using similar methods to those used here, and that the resulting models would have an analogous relationship to the corresponding quantum double models.
			
			It may also be possible to pursue a similar generalization of the color codes based on fusion categories rather than finite groups, in the spirit of the string-net models~\cite{LevinSN05}, of which the quantum double models are examples~\cite{Buerschaper2009}.

%------------------------------------------------------------------------------------------------------------%

\begin{acknowledgments}
	We thank Jonas Anderson, Stephen Bartlett, Hector Bombin, Chris Cesare, Andrew Doherty, Leander Fiedler, and Steve Flammia for fruitful discussions and helpful comments. This work is supported by the ARC via the Centre of Excellence in Engineered Quantum Systems (EQuS) project number CE110001013, by the ERC grant QFTCMPS, and by the cluster of excellence EXC 201 Quantum Engineering and Space-Time Research.
\end{acknowledgments}

%------------------------------------------------------------------------------------------------------------%

\end{document}